# Ultrasensitive interplay between ferromagnetism and superconductivity in NbGd composite thin films

*Ambika Bawa[1], Anurag Gupta[1], Sandeep Singh[2], V.P.S. Awana[1] & Sangeeta Sahoo[1]\**

*[1]Quantum Phenomena & Applications, CSIR-National Physical Laboratory, Council of Scientific and Industrial Research, Dr. K. S. Krishnan Marg, New Delhi, India- 110012*

*[2]Sophisticated and Analytical Instrumentation, CSIR-National Physical Laboratory, Council of Scientific and Industrial Research, Dr. K. S. Krishnan Marg, New Delhi, India- 110012*

*\*Correspondence and requests for materials should be addressed to S.S. (sahoos@nplindia.org)*

A model binary hybrid system composed of a randomly distributed rare-earth ferromagnetic (Gd) part embedded in an s-wave superconducting (Nb) matrix is being manufactured to study the interplay between competing superconducting and ferromagnetic order parameters. The normal metallic to superconducting phase transition appears to be very sensitive to the magnetic counterpart and the modulation of the superconducing properties follow closely to the Abrikosov-Gor'kov (AG) theory of magnetic impurity induced pair breaking mechanism. A critical concentration of Gd is obtained for the studied NbGd based composite films (CFs) above which superconductivity disappears. Besides, a magnetic ordering resembling the paramagnetic Meissner effect (PME) appears in DC magnetization measurements at temperatures close to the superconducting transition temperature. The positive magnetization related to the PME emerges upon doping Nb with Gd. The temperature dependent resistance measurements evolve in a similar fashion with the concentration of Gd as that with an external magnetic field and in both the cases, the transition curves accompany several intermediate features indicating the traces of magnetism originated either from Gd or from the external field. Finally, the signatures of magnetism appear evidently in the magnetization and transport measurements for the CFs with very low (< 1 at.%) doping of Gd.

A coupled superconductor (SC) - ferromagnet (FM) hybrid system can provide a model platform to study the interaction between two fundamentally contrasting many body ground state problems in one system where the conflicts appear due to the incompatible nature of the order parameters of its individual components. One of such existing experimentally accessible electronic hybrid systems is SC-FM based composite thin film (CF) with controllable parameters tuned by its composition. For an s-wave SC is in contact with a FM as in a CF, a modulation of the superconducting order parameter is the most pronounced effect which occurs mostly due to the proximity effect at the SC-FM interfaces, exchange interaction and stray fields originated from FM[1]. In addition, varieties of other interesting physical phenomena like coexistence of magnetism and superconductivity[2], enhancement of superconductivity due to magnetic interaction[3,4], domain wall superconductivity[5,6], high critical current for power electronic application[7], significant pinning enhancement due to the integration of FM particles into the SC matrix[8,9], existence of long-range spin-polarized supercurrents[10,11] etc. have been observed in SC-FM based hybrid systems. Besides, being real multicomponent percolating physical systems, CFs also offer to study the fundamental aspects of electronic, transport, and magnetic properties of individual percolating networks and understand the quantum phase transitions [12,13].

To construct the CFs, we employ one of the commonly explored rare-earth magnets, Gd, which has also been used to impact the superconducting properties while coupled to a superconductor[14,15]. As per the SC material is considered, Nb offers the highest critical temperature for an elemental superconductor. Combination of these two would possibly leads to one of the simplest selection of SC-FM based systems. Since 1960s, NbGd based hybrid systems in various geometries have been used to study the interplay between s-wave superconductivity in Nb and ferromagnetism mediated by 4f localized electrons in Gd[8,16-21]. Recently, we have demonstrated that phase slip processes can be triggered by introducing magnetic Gd particles into superconducting Nb matrices[22]. Here, we study the modulation of the superconducting properties together with an investigation for the existence of any characteristic magnetic properties in the limit of very dilute magnetic doping in NbGd based CFs.

The CFs were grown at room temperature with thicknesses in the range between 50 to 110 nm directly on Si (100) substrate. We observe a strong pair breaking effect induced by magnetic Gd and the modulation of the superconducting critical temperature $T_c$ closely follows the AG theory related to the magnetic impurity induced pair breaking mechanism for an s-wave superconductor[23]. Our results clearly demonstrate that Gd is strongly detrimental to restoring the superconductivity in the CFs and a critical concentration ($c_{cr}$) of about 1.3 at.% of Gd can completely destroy the superconductivity. This is in very much contrast to a recently reported result where the authors claimed to have superconductivity sustained for up to ~ 40 at.% of Gd incorporation with a Nb buffer layer (50 nm)[8]. To avoid any contribution from the buffer layer and hence to directly probe the superconducting properties of CFs we do not use any SC buffer layer. Furthermore, the present findings support some of the earlier works reporting no clear superconducting transitions for the CFs having Gd more than about 5 at.%[18,24].

As a true interplay between SC and FM, the signatures of magnetism have been also observed by the paramagnetic Meissner effect (PME)[25-27] in the temperature dependent DC magnetization measurements (M-T) for the CFs while only diamagnetic Meissner effect is observed in pure Nb films. With increasing Gd above the $c_{cr}$, a paramagnetic transition occurs and the related transition temperature $T_m$ behaves linearly with Gd concentration, $c$. Besides, we observe a non-trivial resistive-superconducting transition curve featured with several intermediate structures initiated and modulated by the Gd incorporation in R-T measurements. These distinct features modulate additively with the applied magnetic field. Thus even with a very dilute magnetic doping of less than 1at.%, the characteristics of magnetic properties are evident in the magnetization as well as in the transport measurements. Finally, we have constructed a compositional phase diagram where the region related to the PME is included in addition to the other phases and have established the magnetic phase diagram for varying composition.

# Results:

The X-ray diffraction (XRD) patterns of as-deposited CFs having different compositions are shown in Figure 1(a). Pure Nb film shows a body-centered cubic (bcc) structure with lattice parameter, *a = 3.31 Å*. All the samples with relatively low $c$, the Gd concentration, up to 15% reveal similar bcc patterns. With increasing $c$, the peak positions shift to lower 2θ values indicating an increase of the lattice parameter due to Gd inclusions. Further, for $c \geq 25$ at.%, appearance of broad features, marked with rectangular blocks in Figure 1, leads to either the presence of both cubic (fcc) and hexagonal phases of Gd[8] or could be a signature of a typical amorphous solids as reported earlier[24]. Pure Gd yields a hexagonal close-packed (hcp) structure with lattice parameters, $a_{hcp}$ = 3.64 Å, and $c_{hcp}$ = 5.88 Å. Figure 1(b) presents the variation of the lattice parameter $a$ for the Nb-rich bcc state with $c$ up to 40%. The inclusion of Gd with larger metallic radius (1.87 Å) than that of Nb (1.43 Å) clearly influences and increases the lattice parameter of the Nb-rich bcc phase.

An atomic force microscope (AFM) is used to study the surface topography of the CFs. Figure 1 (c)-(f) represents the morphological evolution of the CFs with increasing $c$. The morphology of a Nb film of about 100 nm thickness is shown in Figure 1(c) while (d)-(f) of Figure 1 represent the same for NbGd films grown in a single batch but having variation in Gd concentration. The changes in the grain size for different $c$-values are evident in the AFM images. The grain size for the Nb samples are about 20-30 nm as reported earlier also[22] whereas, with increasing $c$ the lateral size of the grains becomes larger and the grain boundaries get blunter. Further, it is visible that the grains start to merge and to form clusters type of morphology for higher concentration of Gd [Figure 1 (f)].

The temperature dependent resistance (R-T) measurements are carried out on CF based devices in conventional 4-terminal geometry. The Device geometry and the electrical connections are shown in the inset of Figure 2(b). A set of normalized R-T curves with varying $c$ are shown in Figure 2(a). The resistance is normalized with the normal state resistance measured at 10 K. The superconducting transition temperature $T_c$ is defined as the temperature corresponding to the maximum value of dR/dT and is shown by the dotted black vertical line for the device with $c = 0.3$ (the red curve). The black open squares in Figure 2(a) present the R-T characteristic for a pure Nb film which shows a sharp transition with an extent of about 0.1 K in temperature and with a $T_c \sim 8.85$ K. With increasing $c$, the normal metal (NM) – superconductor transition curves move towards lower temperature. At $c = 0.95$, the resistance reduced to ~1 % of its normal state value. Further, the transitions get broader and hence an increase in transition width with increasing $c$ as manifested in the R-T measurements. For $c \geq 1$, the devices do not show a complete zero-resistance transition from its normal state down to the temperature limit of the measurement setup of ~ 2 K. Within this narrow range of Gd concentration we observe a crossover between superconducting to non-superconducting states as well as an obvious modulation of the superconducting transition width mediated by the magnetic contribution in the CFs.

The suppression of the $T_c$ is consistent with the pair breaking picture introduced by Abrikosov and Gor'kov (AG) for an s-wave superconductor in the presence of magnetic impurities. As described by the AG theory[23], the normalized $T_c$ can be denoted as:

$$ln\left(\frac{T_{c0}}{T_c}\right) = \psi\left(\frac{1}{2} + \frac{\alpha T_{c0}}{2\pi T_c}\right) - \psi\left(\frac{1}{2}\right) \qquad (1)$$

where, $T_{c0}$ is the critical temperature of pure Nb film, $\alpha = \frac{\hbar}{2\tau k_B T_{c0}}$ is the pair breaking parameter with $\tau$ as the quasi-particle lifetime controlled by the scattering from magnetic impurities. Assuming isotropic scattering in the limit of very low impurity, $\tau$ can be related to the impurity concentration $c$ as[28,29], $\frac{1}{\tau} \propto c$ where $c$ is in at.%. $\psi$ is the digamma function and can be expressed as,

$$\psi\left(\frac{1}{2} + x\right) = \psi\left(\frac{1}{2}\right) + \frac{\pi^2 x}{2} + O(x^2) \qquad (2)$$

In the limit of very dilute doping with $\hbar/\tau \ll k_B T_c$, equation (1) can be rewritten as,

$$\frac{T_c}{T_{c0}} = \left[1 - \left(\frac{\alpha\pi}{4}\right) - \left(\frac{\alpha\pi}{4}\right)^2\right] \qquad (3)$$

Now from equation (3) if we neglect the quadratic term for a very low range of impurity concentration, a linear dependence is expected for the critical temperature to the impurity concentration. From equation (1) we note that for a critical value of $\alpha$, $\alpha_{cr}$, $T_c$ vanishes. This implies that the superconductivity can sustain up to a critical impurity concentration $c_{cr}$.

To envisage experimentally the dependence of the $T_c$ on $c$, we have extracted $T_c$ values from each R-T data for ~30 samples varying in $c$. Each $T_c$ is normalized by $T_{c0}$, the critical temperature of Nb and the normalized $T_c$ is plotted against $c$ in Figure 2 (b). The black squares represent the experimental results whereas the red solid curve is a fit using equation (3). The black stars represent the $T_c$ of measured samples showing partial/no transition down to 2 K. For these samples the $T_c$ values are placed at the 0 K line in Figure 2(b). The critical concentration $c_{cr}$ is considered at the base temperature of the measurement setup as indicated in Figure 2(b) and is about 1.2 at.%.

As the Meissner effect is another hallmark of superconductivity we have verified the dependence of $T_c$ on $c$ through DC M-T measurements also. Figure 3(a) displays a set of M-T curves measured on CFs with different $c$ values. The black squares represent the M-T behaviour of a Nb film with a $T_c \sim 8.7$ K as shown by the arrow in Figure 3(a). Here, the $T_c$ is defined as the temperature at which a CF starts to show negative magnetization. The measured $T_c$ values for Nb from both R-T and M-T measurements are close to its bulk value ($T_{c0}^{bulk} = 9.2$ K).

Similar to the R-T measurements, the M-T transition curves show a general trend to shift towards lower temperature with increasing $c$ [Figure 3(a)]. Likewise, we have extracted the $T_c$ values from each M-T curves with different $c$ and the related dependence is shown in Figure 3 (b). Up to a value of $c \sim 1.4\%$, the variation of $T_c$ on $c$ follows an analogous trend as that is observed from R-T measurements in Figure 2(b). A fit using equation (3) shown by the red curve in Figure 3(b) follows closely to the experimental points in this range of $c$. However, above $c \sim 1.4\%$, the diamagnetic transitions (black circles) occur irregularly for some of the samples and for the rest we observe no transition (black stars) and the related modulation in $T_c$ with $c$ is shown by the dotted curve in Figure 3(b). A representative M-T curve is shown in the inset of Figure 3(b) with $c = 4.7\%$ which is marked with pink dotted circular region in the main panel. Here, we observe a clear diamagnetic transition at $\sim 6.2$ K. However, the amplitude of the magnetization is $\sim 50$ times smaller than that for the samples presented in Figure 3(a). This might be a weak transition originated locally through any percolated path but not through the whole sample and can be treated as an incomplete transition. Incomplete/partial transitions are observed in R-T measurements (Figure S3 in the supporting material) also for the devices with higher values of $c$.

It is apparent from the M-T and R-T measurements that the superconducting $T_c$ collapses rapidly to the base temperature limit with increasing $c$ from 0 to $\sim 1.3$ %. The close agreement between experimental results and the fit using the AG theory indicates the pair breaking mechanism induced by the magnetic doping as the origin of this suppression of superconductivity. The critical concentration $c_{cr}$ of Gd provides a crossover from superconducting to normal state in these CFs. Indeed, the values of $c_{cr}$ from the R-T ($\sim 1.2\%$ at 2 K) and M-T measurements ($\sim 1.4\%$ at 1.8 K) are very close.

It is clear that the magnetic doping is detrimental for the superconducting properties in the CFs and the $T_c$ gets degraded with increased magnetic constituents. However, the detail mechanism behind the degradation of the superconductivity might be understood from the evolution of the SC-NM transition with the concentration of the FM part. To obtain an insight into the transition region of the R-T characteristics we display the R-T data shown in Figure 2(a) in a semi-logarithmic scale in Figure 4(a). As seen earlier, the Nb device adheres a sharp transition down to its complete superconducting state. In contrast, NbGd devices expose some distinct features described by kink/shoulder type of broadenings which get prominent with increasing $c$. The colored arrows in Figure 4(a) indicate the formation of the features along with their evolution with increasing $c$. Hence, the question arises whether the featured kinks in the NbGd devices imply any trace of magnetism appearing from the Gd content.

Further, we emphasize a closer look onto the M-T data measured in both the field cooled (FC) and zero field cooled (ZFC) conditions for a few representative samples presented separately in Figure 4(b). For clarity, we shifted the M-T curves in the y-direction with the order of increasing $c$ values and the dotted baselines attached to each individual curve represent the zero-magnetization state obtained by subtracting the background magnetization. The bottom black circles represent the M-T measurements for a Nb sample with a $T_c \sim 8.7$ K. The similar behaviour is observed for the next sample (the red circles) while moving upward in Figure 4(b) except for a very little fluctuation in the related FC data close to its $T_c$. Moving to the next sample in upward direction with a little higher $c$ value (blue circles), we witness a very different behaviour in ZFC M-T characteristic which shows a crossover from a positive to negative magnetization while decreasing the temperature. Unlike the former two samples, the FC data for this sample shows a strong positive magnetization and the temperature $T_m$ for the onset of positive magnetization appears to be the same at which the positive magnetization appears for the ZFC M-T data. This scenario is very similar to the paramagnetic Meissner effect (PME) observed in earlier studies[30-33]. We highlight the region of paramagnetic to diamagnetic transition by the orange dotted rectangular region in Figure 4(b). With increasing $c$ the PME is still observed for next couple of samples in Figure 4(b), however, the amplitude of the paramagnetic signal gets reduced and the PME disappears for the sample with $c = 1.3\%$. From the M-T measurements with further increasing $c$, we mostly observe an onset of a positive magnetization and the related temperature $T_m$ increases linearly with $c$. We note that due to a smaller paramagnetic signal, $T_m$ values are not distinguishable from the background in the same scale for all the individual samples presented in Figure 4(b). However, their individual data clearly exhibit the transition to paramagnetic state over the background signal (see Figure S4 in the supporting material).

There are two most common and well accepted arguments behind the PME, namely, the d-wave symmetry of quasiparticle pairing and the flux compression[25]. The former is not relevant in our case since we observe a strong suppression of $T_c$ with the concentration of magnetic impurities indicating the SC is of s-wave symmetry as described in the AG theory[23]. The latter, i.e., the magnetic flux compression is possible due to the inhomogeneity present in these CFs[34]. The PME effect has been observed in SC-FM based hybrid systems where the effect is attributed to the flux compression due to spontaneous vortex formation[35] and also to the interplay between SC and FM in the hybrids[36]. Recently, we also have shown that the magnetic doping may lead to the vortex formation[22]. Further, according to the Ginzburg-Landau (GL) theory, the presence of vortices inside a thin superconducting sample can cause the PME by the compression of the flux trapped inside the vortices[37,38]. The kink types of features in R-T characteristics [Figure 4(a)] can be of support to the picture of vortex formation[39] and hence the flux compression. However, for better understanding of the origin of the PME one needs to study the field dependence magnetization behaviour which is out of the scope of this study. Here, we emphasize that the traces of magnetism are evident in both the DC magnetization and the R-T measurements for these NbGd based CFs with less than 1 at.% of Gd in it.

For further support in favor of the signatures of the magnetic characteristics we have performed the R-T measurements under an external magnetic field applied perpendicular to the plane of the samples and their evolution with the B-field is presented in Figure 5. Figure 5(a) shows a set of field dependent R-T measurements data for a Nb device. We observe a sharp transition from the resistive state to its superconducting state under low applied field with a reduced $T_c$ for increased B-field. At relatively higher B-field, the sharpness of the transition gets distorted and it becomes wider. For example at B = 500 mT, the transition curve accompanies a kink type of structure (shown by an arrow) indicating an intermediate state possibly due to the vortex formation under the magnetic field for a type-II superconductor[39]. Thus the applied B-field above a threshold value causes similar types of features/kinks in the R-T characteristics for the Nb film as they appeared for the NbGd samples under no external field. The Field dependent R-T measurements are being carried out for NbGd composite devices also and the results for a few illustrative devices with $c$ = 0.3, 0.5, 0.65 are presented in Figures 5 (b), (c), and (d), respectively. In Figure 5(b) with $c$ = 0.3 and for the fields up to 100 mT, the transition curves look broader compared to that of the Nb device. However, this device with an applied field of about 200 mT facilitates the shouldering effect which was absent under no external field. At B = 200 mT, a small kink starts to appear in the transition characteristic and it gets prominent with increasing the field. Similar type of features appears for devices with higher values of $c$ [Figure 4(a)] under no external field as clearly manifested in Figures 5 (c) and (d). In addition to the broadening of the transition width with increasing c, the shouldering features are getting wider and stronger and they are observed to modulate in a similar fashion with the external field.

It is apparent from Figure 4(a) and Figure 5 that the kinks/ shouldering features appeared in the R-T characteristics originate either from the magnetic particles embedded in the CF or from the external magnetic field suggesting their origin to be related with the magnetism. The features start to appear at higher field for the devices with lower $c$ and *vice versa*. This is a clear indication that the applied field and the concentration of the magnetic counterpart in the CFs supplement each other. Hence it is straightforward that the magnetic component of the CFs generates an effective magnetic field which probably is originated from the exchange interaction and/or the stray field associated with the magnetic particles[39]. The magnetic contribution of Gd in the CFs is also evident by the appearance of positive magnetization in ZFC and the occurrence of the PME in the M-T measurements as shown in Figure 4(b). Further, Nb is a type-II superconductor and the vortex states are expected to play an important role while studying the interaction between the superconductivity and the ferromagnetism in Nb-Gd based hybrid systems. The incorporation of magnetic Gd into the Nb matrix can lead to vortex state due to its internal field and also it can act as the pinning centers under an applied magnetic field. The shouldering type of features appearing in the R-T curves open up a possibility to have spontaneous vortices in the CFs having the Gd concentration above a threshold value. In this study, the features start to appear for $c = 0.5\%$ which could be the threshold value of $c$ to observe the features at zero external B-field.

## Discussion:

Based on the detailed experimental investigations, we have formulated (i) a compositional phase diagram and (ii) a magnetic phase diagram for selected compositions. The dependence of the characteristic transition temperatures on the concentration of magnetic impurities leads to the compositional phase diagram as shown in Figure 6(a). The black squares and the green circles represent the superconducting critical temperatures measured through R-T and M-T measurements, respectively. The red solid curve (in the left), representing the best fit to the experimental points using the AG theory, illustrates the phase boundary of the SC state (the shaded area in dark green). Above a certain minimum value of the Gd concentration, both the FC and ZFC M-T measurements show a positive magnetization indicating a paramagnetic phase of the CFs under consideration. The paramagnetic transition temperature $T_m$ is the temperature related to the onset of the positive magnetization. The $T_m$ is observed to depend linearly to the concentration of the magnetic component and the purple spheres in Figure 6(a) represent the experimental values of $T_m$ while the attached red curve in right is a linear fit. The region below this line in the phase diagram indicates the PM phase. However, For a few NbGd samples with $c$-values in the range of $0<c<c_{cr}$, a crossover from paramagnetic to diamagnetic state has been observed in the ZFC M-T measurements while their M-T characteristics in FC condition show positive magnetization down to the measurement limit of 1.8 K [Figure 4 (b)]. This phenomenon is known as the PME. The yellow region, between the SC and the PM states illustrates the region related to the PME. The rest part of the phase diagram obviously represents the NM phase of the CFs. This is noteworthy to mention that the electronic states are very sensitive to the magnetic impurity as revealed in the phase diagram through the appearance of the PME region.

For constructing the magnetic phase diagram we have extracted the $T_c$ values from the R-T measurements performed under externally applied magnetic field as previously shown in Figure 5. For each device with particular $c$-values, we map the B-$T_c$ phase diagram. Figure 6(b) contains a set of B-$T_c$ phase diagram for a few representative samples with different *c*-values. The shaded areas represent the SC state while the solid lines connect the experimental values. The dotted lines are the extrapolations to the base temperature, 2K. We have used a gradient in the color for the shaded area to indicate the inclusion and variation of the magnetic constituent. The region of superconducting area in the phase diagram decreases with increasing *c* which is consistent with the pair breaking mechanism discussed previously. The experimental data along with their linear fits for the same set of devices are shown in Figure 6(c). From the GL theory near the $T_c$, a linear dependence of $T_c$ on the B-field is expected for a dirty superconductor[40,41]. We have extracted the $B_{c2}(T^*)$ value for each device at $T^* = T/T_c = 0.75$ from the linear fit and its dependence on *c* is displayed in the left axis of Figure 6 (d). As expected, $B_{c2}(T^*)$ decreases with increasing *c* since the concentration of the magnetic part and the external magnetic field behave in a complimentary fashion. From the GL theory the coherence length $\xi_{GL}(0)$ can be obtained by the slope of the linear fit of the B-$T_c$ data as[42],

$$\xi_{GL}(0) = \left[\frac{\phi_0}{2\pi T_c (dB_{c2}/dT)_{T_c}}\right]^{1/2} \tag{4}$$

Where, $\phi_0$ is the flux quantum. We have calculated the values of $\xi_{GL}(0)$ for different Gd concentration in the range between 0 to 0.8 % and the related variation is shown by the red stars in Figure 6(d). $\xi_{GL}(0)$ increases with *c* and this is consistent with the dependence of $B_{c2}$ on *c*. However, with increasing *c* the mean free path gets reduced due to increased scattering and hence the coherence length should also be reduced[42]. Also note that the $\xi_{GL}(0)$ for the Nb sample is much lower than that of a pure bulk Nb. This is because the Nb films were grown in the same chamber where NbGd samples were prepared and it is quite likely that our Nb films are in the dirty limit with a low residual resistivity ($R_{300 K}/R_{10 K} \sim 2.78$). Another reason could be the grain size of the Nb films which are in the range of 20-30 nm[22]. The similar value of the $\xi_{GL}(0)$ has been reported for Nb with grain sizes ~20 nm[43]. Further, the enhancement of the $\xi_{GL}(0)$ with Gd concentration is not only related to the grain size since at this low range of Gd concentration the variation in the grain sizes are not detectable from the AFM studies. The enhancement can be related to the interaction between SC and FM and further study is needed for clear understanding.

Finally, we have demonstrated a pronounced and controlled modulation of the superconducting properties in NbGd based CFs for Gd concentration in the range between 0 to 1.4 at.% by performing M-T and R-T measurements. A critical concentration of Gd ~ 1.3 at.% is found above which superconductivity disappears. Besides, the traces of magnetism with less than 1 at.% Gd in the CFs have been appeared evidently through the PME in M-T measurements and the shouldering effect in R-T measurements. However we do observe a non-monotonous oscillatory dependence of $T_c$ on $c$ from M-T measurements for the samples with $c \geq 1.4$ at.%. With high $c$-value, diamagnetic transition can occur partially through any percolating path or due to in-homogeneities mediated by possible Gd clustering. Also the AFM images in Figure 1(d)-(f) indicate of possible Gd clustering at high Gd concentration. There could be other effects like the formation of the triplet states originating from the interaction between SC and FM which could lead to an oscillatory behaviour in the $T_c$ as reported earlier[20]. Nevertheless, the R-T measurements did not show similar behaviour except for the partial transitions (Figure S3 in the supporting material). Hence we can assume that the non-monotonous behaviour is related to the partial transition.

## Methods:

We fabricated NbGd based CFs by co-sputtering of Gd (99.95%) and Nb (99.99%) using an ultra-high vacuum (UHV) DC magnetron sputtering system. CFs are grown on Si (100) substrate with a top layer of 300 nm thick thermally oxidized $SiO_2$ as the dielectric. The sputtering chamber was evacuated to less than $5 \times 10^{-9}$ Torr before each deposition and the co-sputtering was performed in an Ar (99.9999% purity) environment at about $3 \times 10^{-3}$ mBar. The deposition rates for Nb and Gd were 3 Å/sec and 1-2 Å/sec, respectively. Apart from the determination of the sputtering rate, we had performed an extensive optimization process with respect to the relative position of the samples from the sputtering targets and the open area for the shutter of Gd target in order to obtain a gradient in the Gd concentration with respect to the relative position of the samples from the targets. Using this technique we have been able to manufacture samples with varying Gd concentration but keeping the Nb part unaltered from one single sputtering run. For different deposition runs we tried to maintain the relevant parameters very closely so that the results could be comparable from samples fabricated in different sputtering runs. On this context we prepared 2 batches with thicknesses ~ 60 nm and ~110 nm, respectively, and each batch contained 6 devices with variation in Gd concentration. Devices from both the batches follow the AG theory independently and the related $c_{cr}$ values remained very close (see the supporting material). This confirms that the devices prepared from different batches with similar conditions can be comparable.

We mainly prepared two types of samples, namely, CFs with dimensions 3.5 mm x 3.5 mm x 100 nm for DC M-T measurements and NbGd based patterned devices for transport studies. CFs, studied in this report, were grown at room temperature with thickness varying in the range of 50-110 nm. A Si capping layer of about 10 nm thickness was sputtered on top of the CFs for both types of samples to avoid any oxidation while exposed to the atmosphere. We employed shadow mask to fabricate multi-terminal devices for conventional 4-terminal transport measurements. Current and voltage leads were designed with Au (100nm)/Ti (20nm) layers followed by the fabrication of about 100 micron wide and about 1300 micron long channel of NbGd based CFs. Careful measurements were done for the estimation of the concentration by energy dispersive spectroscopy (EDS) analysis using a field emission scanning electron microscope by Zeiss. For each sample, EDS measurements were performed at 5-10 different places and the average value was considered for the analysis along with the uncertainty range estimated from the measured composition at different places. Structural characterization of CFs with varying $c$ was done by grazing incidence X-ray diffraction (GIXRD) using a θ-2θ x-ray diffractometer (Philips X'pert pro X-ray diffractometer) with Cu-k$_\alpha$ radiation operating at 40 kV and 20 mA and the morphological studies were performed by an atomic force microscopy (AFM). A SQUID magnetometer from Quantum Design was used to perform the M-T measurements. The transport measurements were carried out in a Physical Properties Measurement System (PPMS) by Quantum Design. The devices were mounted on a puck and soldered for the electrical measurements.

# References:

1    Buzdin, A. I. Proximity effects in superconductor-ferromagnet heterostructures. *Rev. Mod. Phys.* **77**, 935 (2005).

2    Aladyshkin, A. Y., Silhanek, A. V., Gillijns, W. & Moshchalkov, V. V. Nucleation of superconductivity and vortex matter in superconductor-ferromagnet hybrids. *Supercond. Sci. Technol.* **22**, 053001 (2009).

3    Wei, T. C., Pekker, D., Rogachev, A., Bezryadin, A. & Goldbart, P. M. Enhancing superconductivity: Magnetic impurities and their quenching by magnetic fields. *Europhys. Lett.* **75**, 943 (2006).

4    Stamopoulos, D., Pissas, M., Karanasos, V., Niarchos, D. & Panagiotopoulos, I. Influence of randomly distributed magnetic nanoparticles on surface superconductivity in Nb films. *Phys. Rev. B* **70**, 054512 (2004).

5    Yang, Z., Lange, M., Volodin, A., Szymczak, R. & Moshchalkov, V. V. Domain-wall superconductivity in superconductor-ferromagnet hybrids. *Nat. Mater.* **3**, 793 (2004).

6    Linder, J. & Halterman, K. Superconducting spintronics with magnetic domain walls. *Phys. Rev. B* **90**, 104502 (2014).

7    Stamopoulos, D., Manios, E., Pissas, M. & Niarchos, D. Modulation of the properties of a low-$T_c$ superconductor by anisotropic ferromagnetic particles. *Physica C* **437-38**, 289 (2006).

8    Palau, A. *et al.* Hysteretic Vortex Pinning in Superconductor-Ferromagnet Nanocomposites. *Phys. Rev. Lett.* **98**, 117003 (2007).

9    Togoulev, P. N., Suleimanov, N. M. & Conder, K. Pinning enhancement in $MgB_2$-magnetic particles composites. *Physica C* **450**, 45 (2006).

10  Bergeret, F. S., Volkov, A. F. & Efetov, K. B. Odd triplet superconductivity and related phenomena in superconductor-ferromagnet structures. *Rev. Mod. Phys.* **77**, 1321 (2005).

11  Eschrig, M. Spin-polarized supercurrents for spintronics. *Phys. Today* **64**, 43-49 (2011).

12  Strelniker, Y. M., Frydman, A. & Havlin, S. Percolation model for the superconductor-insulator transition in granular films. *Phys. Rev. B* **76**, 224528 (2007).

13  Liu, X., Panguluri, R. P., Huang, Z.-F. & Nadgorny, B. Double Percolation Transition in Superconductor-Ferromagnet Nanocomposites. *Phys. Rev. Lett.* **104**, 035701 (2010).

14  Matthias, B. T., Suhl, H. & Corenzwit, E. Spin Exchange in Superconductors. *Phys. Rev. Lett.* **1**, 92 (1958).

15  Kim, H. *et al.* Effect of magnetic Gd impurities on the superconducting state of amorphous Mo-Ge thin films with different thickness and morphology. *Phys. Rev. B* **86**, 024518 (2012).

16  Strunk, C., Sürgers, C., Paschen, U. & Löhneysen, H. v. Superconductivity in layered Nb/Gd films. *Phys. Rev. B* **49**, 4053 (1994).

17  Koch, C. C. & Love, G. R. Superconductivity in Niobium Containing Ferromagnetic Gadolinium or Paramagnetic Yttrium Dispersions. *J. Appl. Phys.* **40**, 3582 (1969).

18  Scholten, P. D. & Moulton, W. G. Effect of ion-implanted Gd on the superconducting properties of thin Nb films. *Phys. Rev. B* **15**, 1318 (1977).

19  Strunk, C., Paschen, U., Surgers, C. & Vonlohneysen, H. Pair-breaking mechanisms in Nb/Gd/Nb films. *Physica B* **194**, 2403 (1994).

20  Jiang, J. S., Davidović, D., Reich, D. H. & Chien, C. L. Oscillatory Superconducting Transition Temperature in Nb/Gd Multilayers. *Phys. Rev. Lett.* **74**, 314 (1995).

21  Strunk, C., Surgers, C., Rohberg, K. & Vonlohneysen, H. Transition-temperature and critical fields of Nb/Gd layers *Physica B* **194**, 2405 (1994).

22  Bawa, A., Jha, R. & Sahoo, S. Tailoring phase slip events through magnetic doping in superconductor-ferromagnet composite films. *Sci. Rep.* **5**, 13459 (2015).

23  Abrikosov, A. A. & Gor'kov, L. P. Contribution to the theory of superconducting alloys with paramagnetic impurities. *Sov. Phys. JETP* **12**, 1254 (1961).

24  Sommer, R. L., Xiao, J. Q. & Chien, C. L. Magnetic and magneto-transport properties of metastable $Gd_xNb_{1-x}$ alloys. *IEEE Trans. Magn.* **34**, 1135 (1998).

25  Li, M. S. Paramagnetic Meissner effect and related dynamical phenomena. *Phys. Rep.* **376**, 133 (2003).

26  Braunisch, W. *et al.* Paramagnetic Meissner effect in high-temperature superconductors. *Phys. Rev. B* **48**, 4030 (1993).

27  Braunisch, W. *et al.* Paramagnetic Meissner effect in Bi high-temperature superconductors. *Phys. Rev. Lett.* **68**, 1908 (1992).

28  Kim, Y.-J. & Overhauser, A. W. Theory of impure superconductors: Anderson versus Abrikosov and Gor'kov. *Phys. Rev. B* **47**, 8025 (1993).

29  Powell, B. J. & McKenzie, R. H. Dependence of the superconducting transition temperature of organic molecular crystals on intrinsically nonmagnetic disorder: A signature of either unconventional superconductivity or the atypical formation of magnetic moments. *Phys. Rev. B* **69**, 024519 (2004).

30  Fang, Y., Yazici, D., White, B. D. & Maple, M. B. Enhancement of superconductivity in $La_{1-x}Sm_xO_{0.5}F_{0.5}BiS_2$. *Phys. Rev. B* **91**, 064510 (2015).

31  Thompson, D. J., Minhaj, M. S. M., Wenger, L. E. & Chen, J. T. Observation of Paramagnetic Meissner Effect in Niobium Disks. *Phys. Rev. Lett.* **75**, 529 (1995).

32  Yuan, S., Ren, L. & Li, F. Paramagnetic Meissner effect in Pb nanowire arrays. *Phys. Rev. B* **69**, 092509 (2004).

33  Papadopoulou, E. L., Nordblad, P., Svedlindh, P., Schöneberger, R. & Gross, R. Magnetic Aging in $Bi_2Sr_2CaCu_2O_8$ Displaying the Paramagnetic Meissner Effect. *Phys. Rev. Lett.* **82**, 173 (1999).

34  Das, P. *et al.* Surface superconductivity, positive field cooled magnetization, and peak-effect phenomenon observed in a spherical single crystal of niobium. *Phys. Rev. B* **78**, 214504 (2008).

35   Xing, Y. T., Micklitz, H., Baggio-Saitovitch, E. & Rappoport, T. G. Controlled switching between paramagnetic and diamagnetic Meissner effects in superconductor-ferromagnet Pb-Co nanocomposites. *Phys. Rev. B* **80**, 224505 (2009).

36   López de la Torre, M. A. *et al.* Paramagnetic Meissner effect in $YBa_2Cu_3O_7/La_{0.7}Ca_{0.3}MnO_3$ superlattices. *Phys. Rev. B* **73**, 052503 (2006).

37   Moshchalkov, V. V., Qiu, X. G. & Bruyndoncx, V. Paramagnetic Meissner effect from the self-consistent solution of the Ginzburg-Landau equations. *Phys. Rev. B* **55**, 11793 (1997).

38   Zharkov, G. F. Paramagnetic Meissner effect in superconductors from self-consistent solution of Ginzburg-Landau equations. *Phys. Rev. B* **63**, 214502 (2001).

39   Xing, Y. T. *et al.* Spontaneous vortex phases in superconductor-ferromagnet Pb-Co nanocomposite films. *Phys. Rev. B* **78**, 224524 (2008).

40   Vicent, J. L., Hillenius, S. J. & Coleman, R. V. Critical-Field Enhancement and Reduced Dimensionality in Superconducting Layer Compounds. *Phys. Rev. Lett.* **44**, 892 (1980).

41   Tinkham, M. *Introduction to Superconductivity*. 2nd edn, (McGraw-Hill, 1996).

42   Orlando, T. P., McNiff, E. J., Foner, S. & Beasley, M. R. Critical fields, Pauli paramagnetic limiting, and material parameters of $Nb_3Sn$ and $V_3Si$. *Phy. Rev. B* **19**, 4545 (1979).

43   Bose, S., Raychaudhuri, P., Banerjee, R. & Ayyub, P. Upper critical field in nanostructured Nb: Competing effects of the reduction in density of states and the mean free path. *Phys. Rev. B* **74**, 224502 (2006).

## Acknowledgments:

We thank Prof. R.C. Budhani for his fruitful discussions and we are thankful to Dr. Ranjana Mehrotra for her support in carrying out this work. We are indebted to Dr. K. K. Maurya for his support in carrying out the GIXRD characterization. The technical help for FESEM imaging, EDS characterization and thickness optimization using the central facilities at CSIR-NPL are highly acknowledged. We are thankful to Mr. M. B. Chhetri for his assistance in building the UHV sputtering unit. We gratefully acknowledge Dr. Sudhir Husale for critical reading of the manuscript and helping us with the invaluable discussions. We are thankful to Mr. Sachin Yadav and Mr. Bikash Gajar for their help and support during the revision of the manuscript. A. B. acknowledges financial support from CSIR-NPL for the research internship. The work was supported by CSIR network project 'AQuaRIUS' (Project No.: PSC 0110).
## Author contributions:

A.B. and S.s.a. designed and fabricated the devices and participated in the measurements. A.G. performed the magnetization measurements in MPMS. S.s.i. conducted the AFM measurements. V.P.S.A. participated in the transport measurements using PPMS. S.s.a. analyzed and interpreted the data and wrote the manuscript. All the authors read and reviewed the manuscript.

## Additional information

Competing financial interests: The authors declare no competing financial interests.

# Figure Captions:

**Figure 1: Structural and morphological characterization.** (a) X-ray diffraction spectra of Nb-Gd composite thin film samples. The curves are shifted for the sake of clarity. The Nb-rich samples show crystalline bcc structure whereas pure Gd film shows hcp structure. Above 25 at.% of Gd a broad peak, indicated by the rectangular block, appears for all the samples. The peaks related to the Nb-rich bcc structures are marked with asterisk ('*') marks. (b) Lattice Parameter, $a$ of the Nb-rich bcc state with Gd concentration $c$ up to 40 at.%. (c) - (f) AFM images showing morphological evolution of 100 nm thick CFs with different Gd content. Surface morphology of (c) Nb ($c = 0$), (d) NbGd ($c = 5\%$), (e) NbGd ($c = 9\%$), and (f) NbGd ($c = 13\%$).

**Figure 2: R-T measurements at zero-field.** (a) Temperature dependent resistance measurements on Nb-Gd based CFs with different Gd concentration, $c$. The resistance is normalized by the normal state resistance measured at 10 K. The superconducting transition temperature $T_c$ is defined as the temperature corresponding to the maximum value in dR/dT and is indicated by the black dashed line for a device with '$c$' = 0.3 ± 0.1. (b) Variation of $T_c$, extracted from R-T measurements, with Gd concentration, $c$. $T_c$ is normalized by $T_{c0}$, the transition temperature of pure Nb film. The scattering points represent the experimental data and the solid red curve is a fit using the Abrikosov-Gorkov (AG) theory. The intersect point between the fit and the base temperature (2 K) provides a critical value of $c$, '$c_{cr}$' representing the crossover from superconducting to normal state transition. The dashed lines close to 2 K represent the temperature limit in the existing measurement setup. Inset: SEM micrograph of a representative device having NbGd as the active channel connected with Au/Ti electrical contacts.

**Figure 3: M-T measurements performed under a perpendicular field of 100 Oe in ZFC condition.** (a) A set of M-T curves for Nb-Gd based CFs with different '$c$' values. The black arrow indicates the $T_{c0}$ value where the Nb film starts to show diamagnetic transition. (b) Variation of $T_c$ with '$c$' from the M-T measurements. $T_c$ is normalized by $T_{c0}$. The scattering points represent the experimental data and the red solid curve represents a fit using the AG theory. The crossing between the fit and the base temperature (1.8 K) provides the critical concentration value '$c_{cr}$' of Gd content. The dashed line close to 1.8 K represents the temperature limit of the measurement setup. The dotted curve is a guide to the eye. Inset: M-T curve of a sample represented by the pink dotted circular region in the main panel of (b).

**Figure 4:** (a) A semi-logarithmic presentation of R-T curves shown in Figure 2(a). The colored arrows indicate the presence of shoulder/kink type of structures appearing for NbGd devices while they are absent in pure Nb device (b) Presentation of selected portion of M-T curves in FC (open circles) and ZFC (solid circles) conditions for a few representing samples close to their transition. For clarity the curves are shifted. The dotted base lines attached to each curve represent the zero- magnetization value. The orange dotted rectangular region demonstrates the paramagnetic Meissner effect (PME) where a positive magnetization is observed just before the diamagnetic transition for ZFC condition while corresponding FC curve remains positive. The superconducting critical temperature $T_c$ and the temperature $T_m$ related to the onset of the positive magnetization from the background are shown by the arrows.

**Figure 5: Evolution of R-T curves with magnetic field applied perpendicular to the sample plane.** Field dependence of R-T characteristics for devices with (a) $c = 0$, i.e. a pure Nb device, (b) $c = 0.3$, (c) $c = 0.5$, (d) $c = 0.65$, respectively. The formation and the positions of the shoulder-type of structures are indicated by the arrows.

**Figure 6: Phase diagram.** (a) A compositional phase diagram constituted by measuring the transition temperatures for devices with different Gd concentration. The $T_c$ values are measured through R-T and DC M-T measurements while the latter also provides the values of magnetic transition temperature $T_m$. (b) The magnetic (B-$T_c$) phase diagram for a few representing samples with different composition. For clarity curves are placed separately along $z$-axis in 3D representation. The solid lines represent the experimental data and the corresponding shaded areas with dark green represent the SC state. The gradient in the color shade indicates the variation in the composition. (c) 2-D representation of (B-$T_c$) phase diagram for the same devices presented in (b). Scattering points represent the experimental values measured from the field dependence of the R-T characteristics from each sample. The solid red lines are the linear fits. (d) Left: Variation of $B_{c2}(T^*)$ values obtained from the linear fit at $T^* = T/T_c = 0.75$ with the Gd concentration. Right: The dependence of the GL coherence length $\xi_{GL}(0)$, calculated from the slope of the linear fit for the B-$T_c$ data, with $c$. The solid lines are guide to the eye.

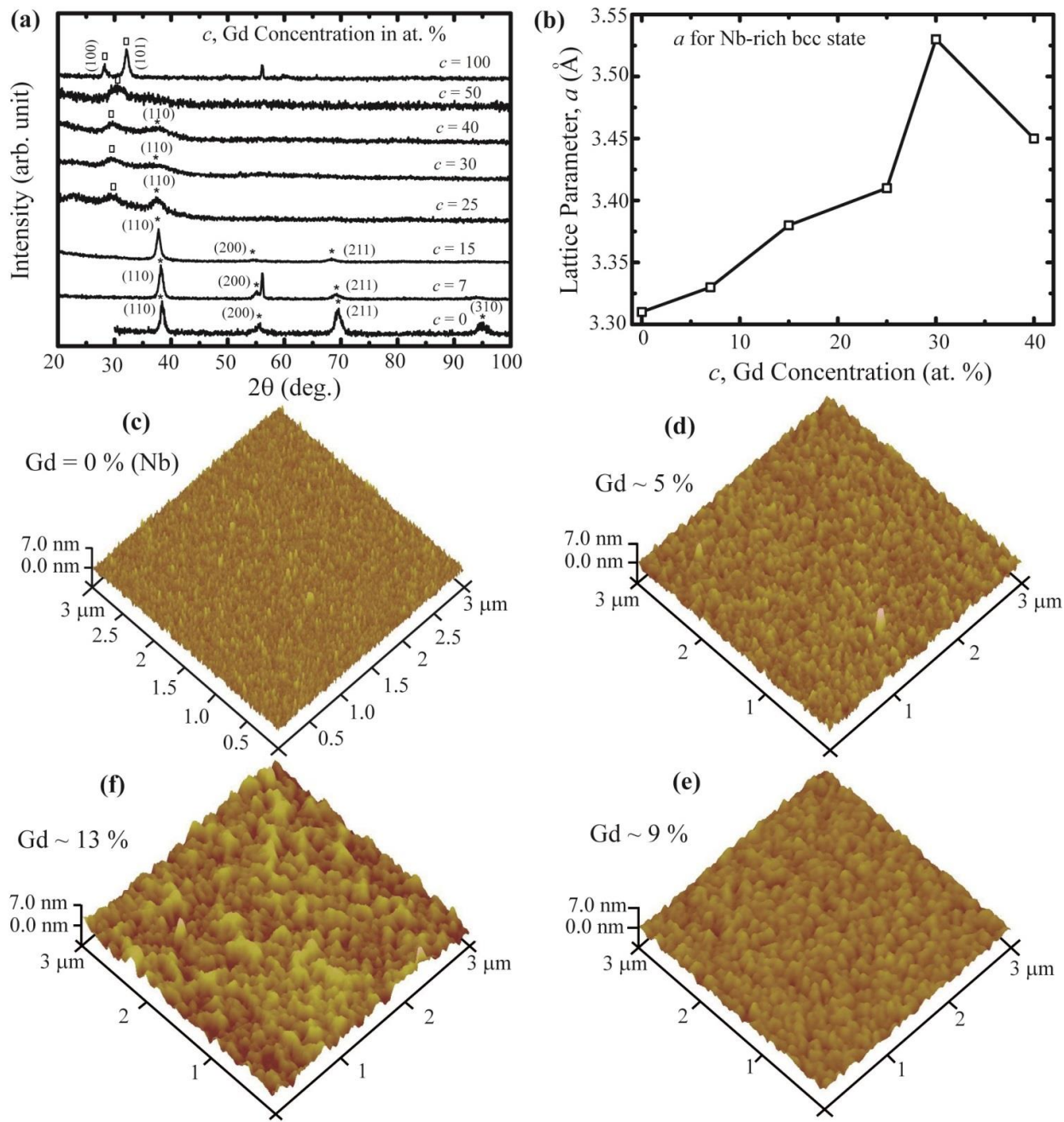

Figure-1

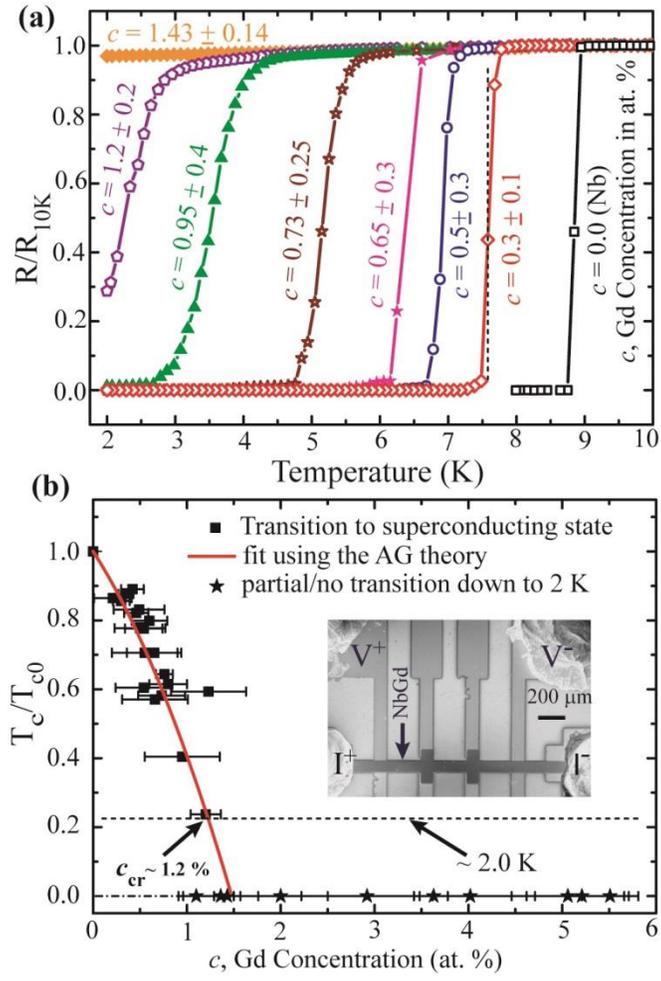

Figure-2

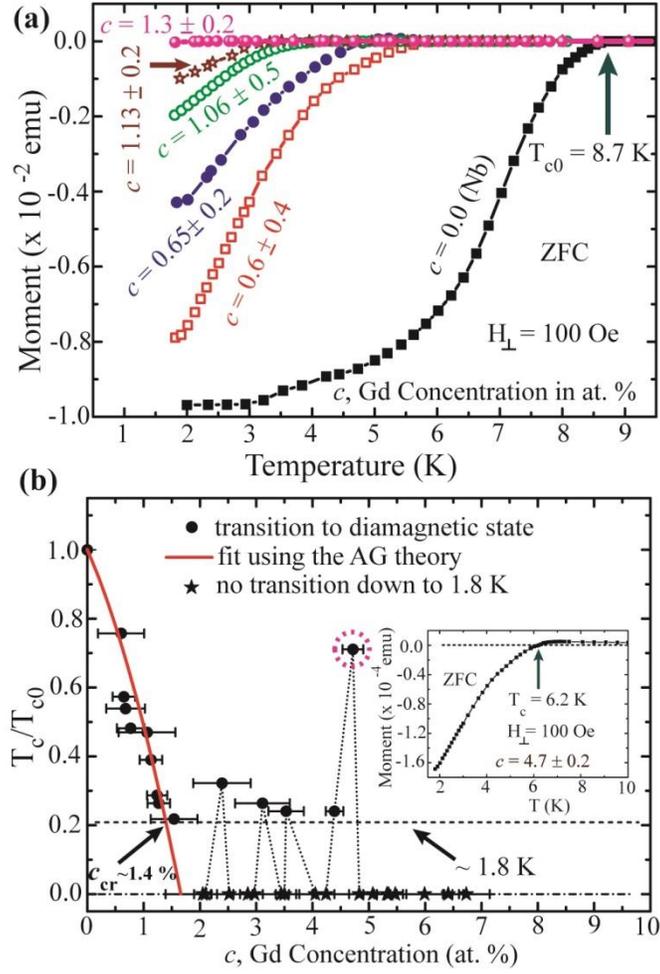

Figure-3

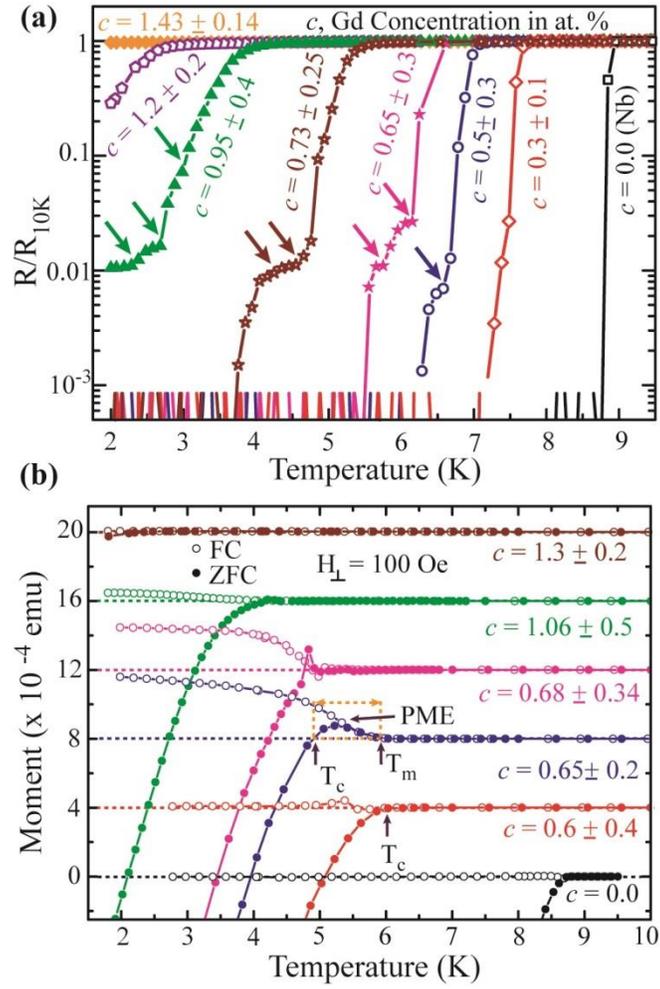

Figure-4

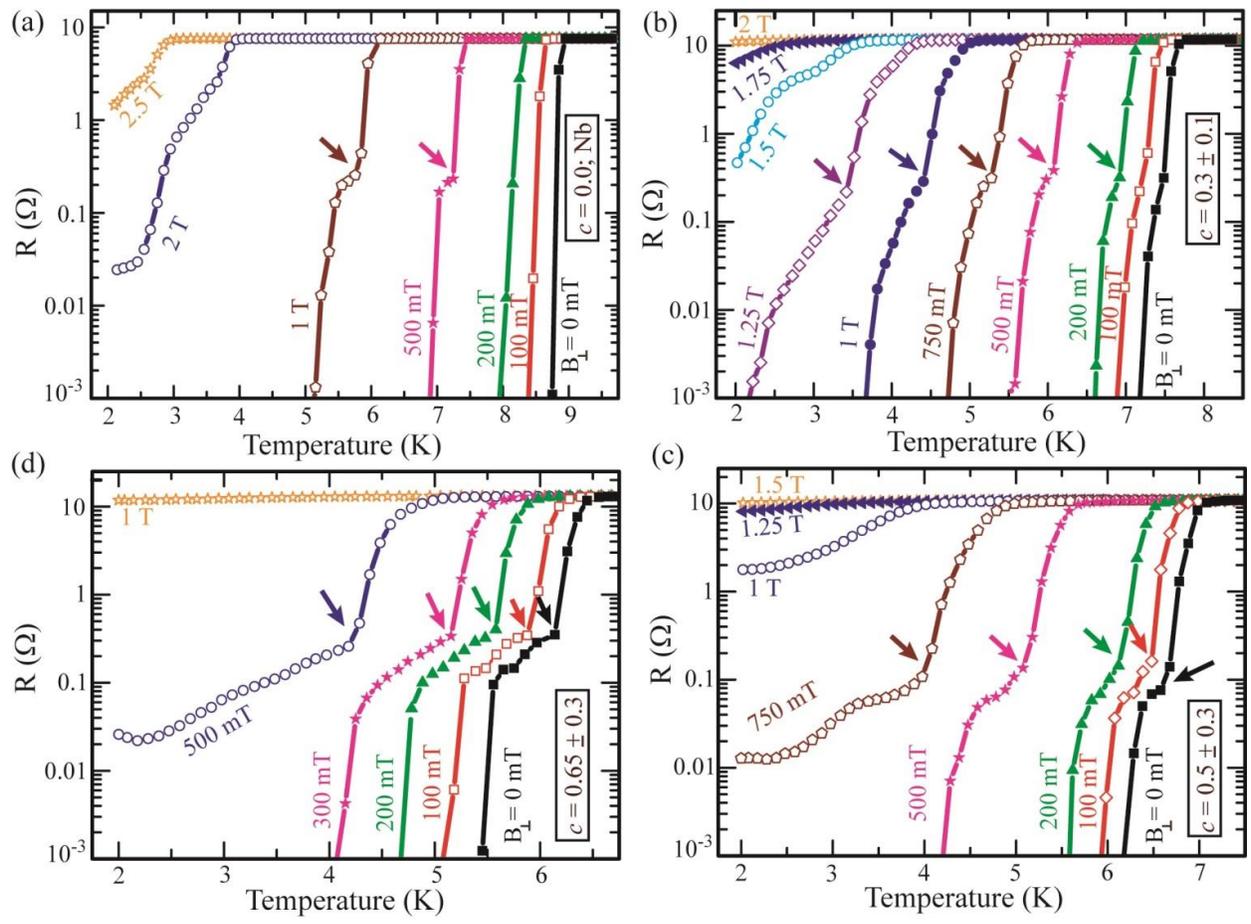

Figure-5

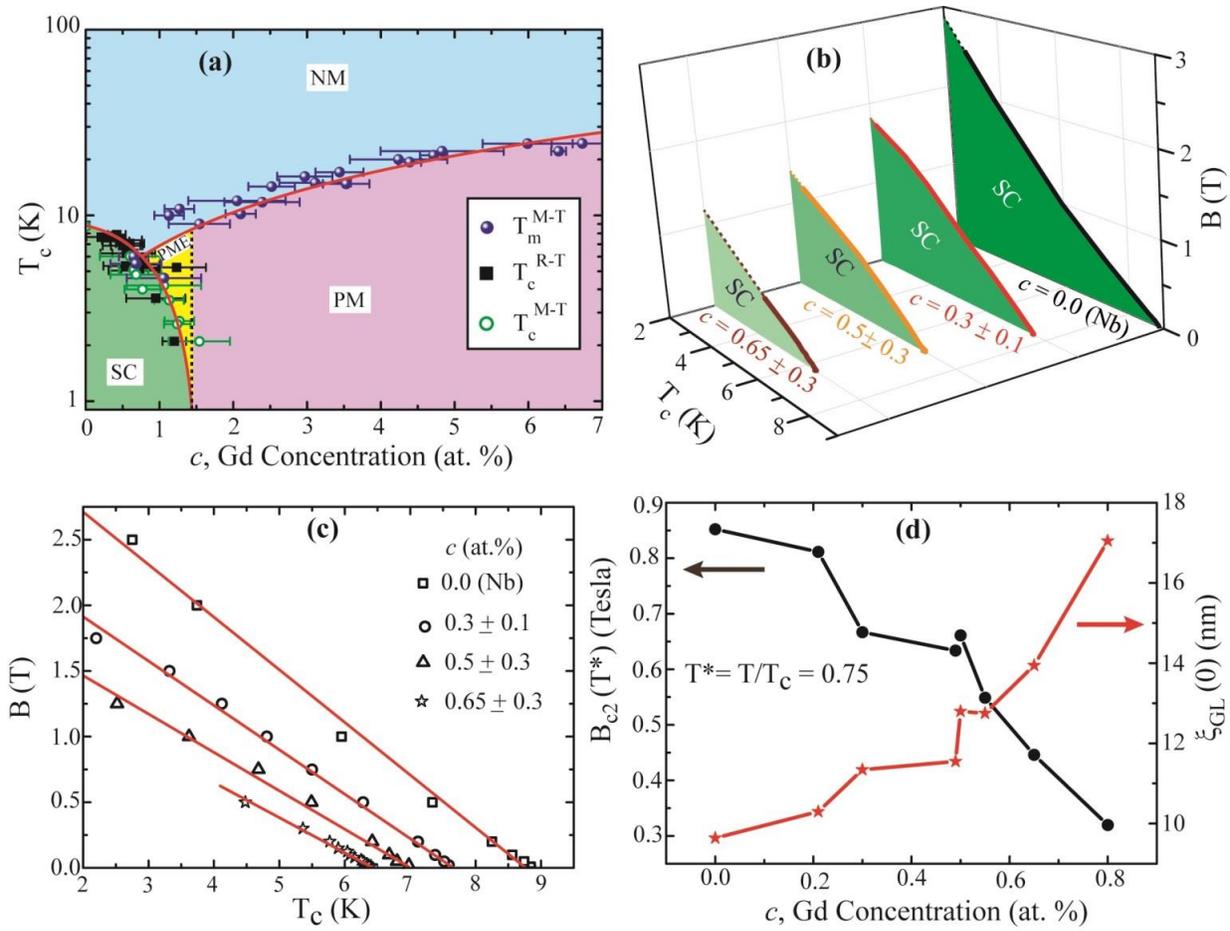

Figure-6

*Supporting Material*

# Ultrasensitive interplay between ferromagnetism and superconductivity in NbGd composite thin films

*Ambika Bawa[1], Anurag Gupta[1], Sandeep Singh[2], V.P.S. Awana[1] & Sangeeta Sahoo[1]\**

[1]*Quantum Phenomena & Applications, CSIR-National Physical Laboratory, Council of Scientific and Industrial Research, Dr. K. S. Krishnan Marg, New Delhi, India- 110012*

[2]*Sophisticated and Analytical Instrumentation, CSIR-National Physical Laboratory, Council of Scientific and Industrial Research, Dr. K. S. Krishnan Marg, New Delhi, India- 110012*

*\*Correspondence and requests for materials should be addressed to S.S.(sahoos@nplindia.org)*

# Fabrication of NbGd based composite films (CFs):

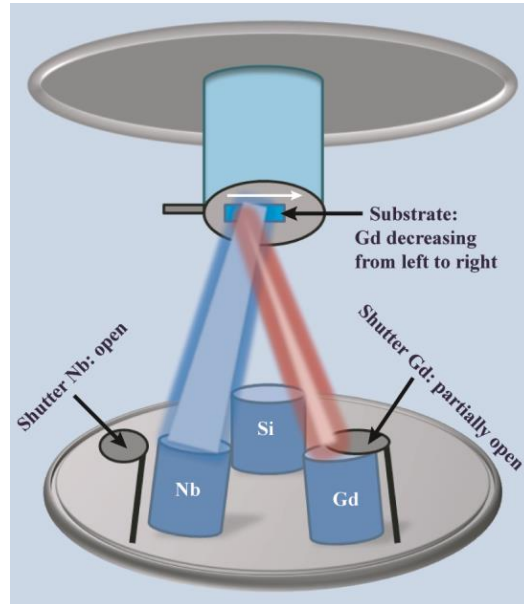

**Figure S1**: Schematic presentation of sample fabrication

We have prepared NbGd based CFs with variations in Gd concentration while keeping Nb concentration almost unaltered throughout the substrate. Figure S1 represents schematically the process during a sputtering run. The sample is placed in such a way that the deposition rate for Nb is uniform for a wide area of the sample while the deposition rate for Gd is varied by the shutter position and the relative position of the sample. The resultant film provides a gradient in Gd concentration along the long side of the strip as indicated by the white arrow. By this way we usually prepare 6 devices at one single run for transport measurement. For the magnetization measurements we use a long Si strip at the beginning and after the sputtering we cut into several pieces of 3.5 mm x 3.5 mm dimensions. We can place more than one strip in parallel and close to each other.

Note that the differences in the $c$-values mentioned in the main manuscript are noticeably small with relatively higher uncertainties which most likely come from the bigger area of the sample with more probable non-uniform distribution of the magnetic particles. However, as mentioned in the experimental section, we paid careful attention towards determining the composition by energy dispersive spectroscopy (EDS) at many places and the relative uncertainties are provided as the error bars accompanied with the average values. Further, the order (increasing and/decreasing) in the *c* values for device to device are consistent with their position during a sputtering run. Also, before the fabrication of the devices used for this study, we had performed an extensive and rigorous optimization process in order to deal with the relative Gd concentration for sample to sample with their specific position inside the sputtering chamber while keeping the Nb concentration almost unaltered.

## Robustness of the sample properties:

In order to compare the samples prepared in different runs we have investigated the modulation of the superconducting properties with Gd concentration for devices made in individual runs and also with variation in the film thickness. Figure S2 represents the variation of the normalized transition temperature with Gd concentration for two sets of devices being fabricated in two separate sputtering runs performed under similar growth conditions. The first set of devices [Figure S2 (a)] are having thicknesses ~60 nm and for the second batch [Figure S2 (a)] the same is about 110 nm. Both the sets independently follow the AG theory as indicated by the red curves [the fits using equation (3) in the main manuscript]. And the critical concentration ($c_{cr}$) values are also close. This clearly states that within this range of thickness values one can have comparable results from the devices grown at different sputtering runs. To get a more accurate value of $c_{cr}$, we have collected the data from many more devices grown at different sputtering runs and the results are presented in the manuscript.

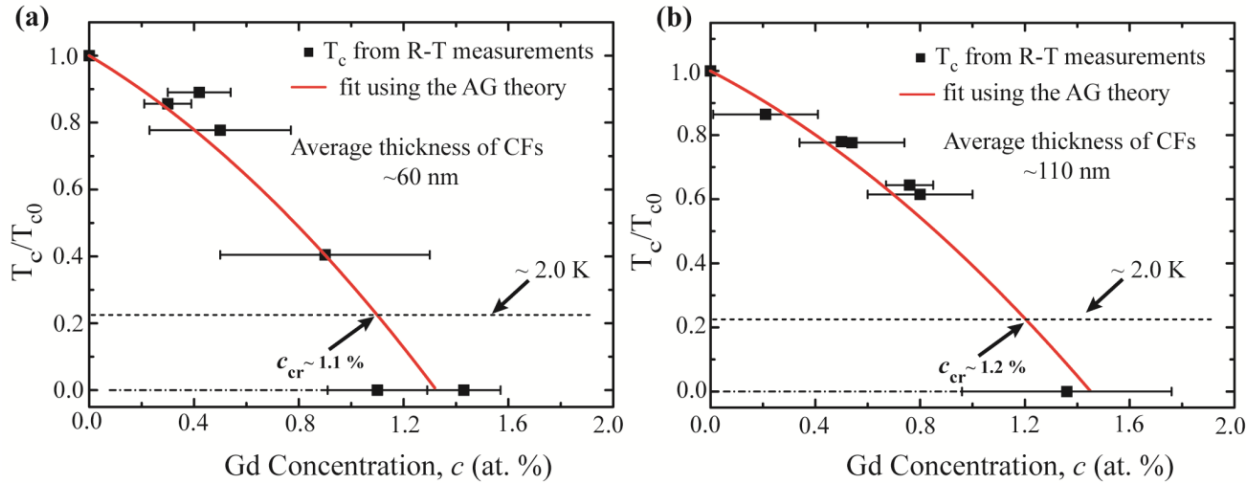

**Figure S2: Variation of $T_c$ with Gd concentration for two different batches of samples.** $T_c/T_{c0}$ vs. *c* for the devices with average thickness (a) ~ 60 nm and (b) ~ 110 nm. The $T_c$ is extracted from R-T measurements. The black squares are the measured values and the red solid curves represent fits using the AG theory [equation (3) in the manuscript]. The dotted lines represent the temperature limit of our measurement setup.

# Partial transition for the devices with higher values of *c* from the R-T measurements:

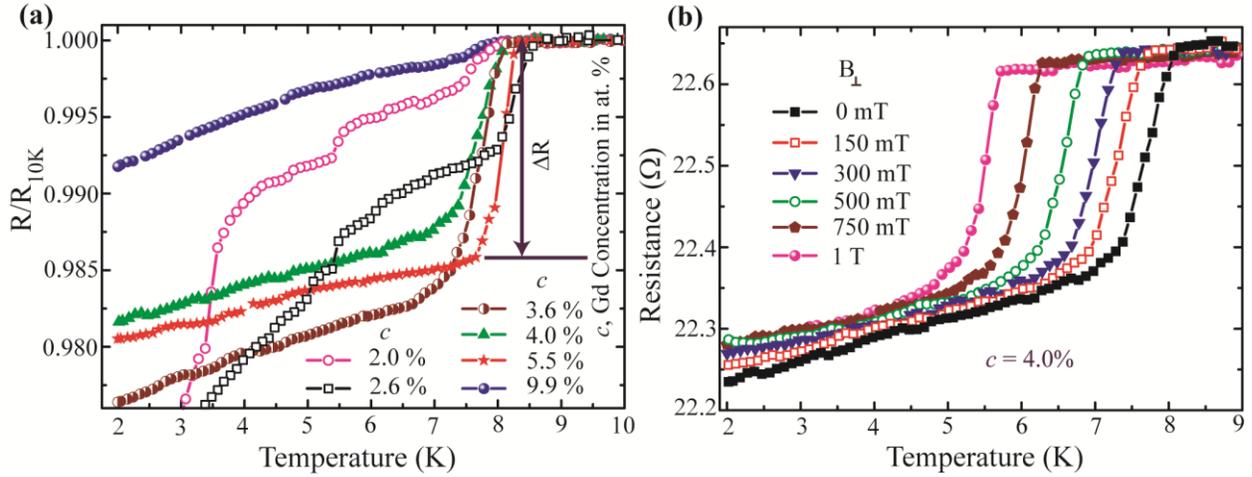

**Figure S3: R-T measurements for the devices with higher Gd concentration.** (a) A set of R-T characteristics at zero external magnetic field. ΔR indicates the resistance drop. (b) The data of R-T measurements performed under an external magnetic field for a representative sample with *c* = 4%

As indicated in the manuscript, we do observe a resistance drop (ΔR) for the samples with high Gd content. In this case, the resistance does not reach its zero value but there is a drop which appears in the R-T characteristics evidently. We call theses transitions as the partial/incomplete transitions and most likely these transitions are local event and not through the bulk of the sample. To understand the origin of these transitions and to address whether they are of superconducting origin, we have measured the R-T dependence under an external magnetic field and we have shown the same in Figure S3(b) for one representative sample (c = 4%). A clear resistance drop appears with a lower temperature shift for higher B-field. This is very similar to what is expected for superconducting transition and hence we call them as the partial/incomplete transitions.

# Variation of the magnetic transition temperature $T_m$ with Gd concentration for samples prepared in a single deposition run:

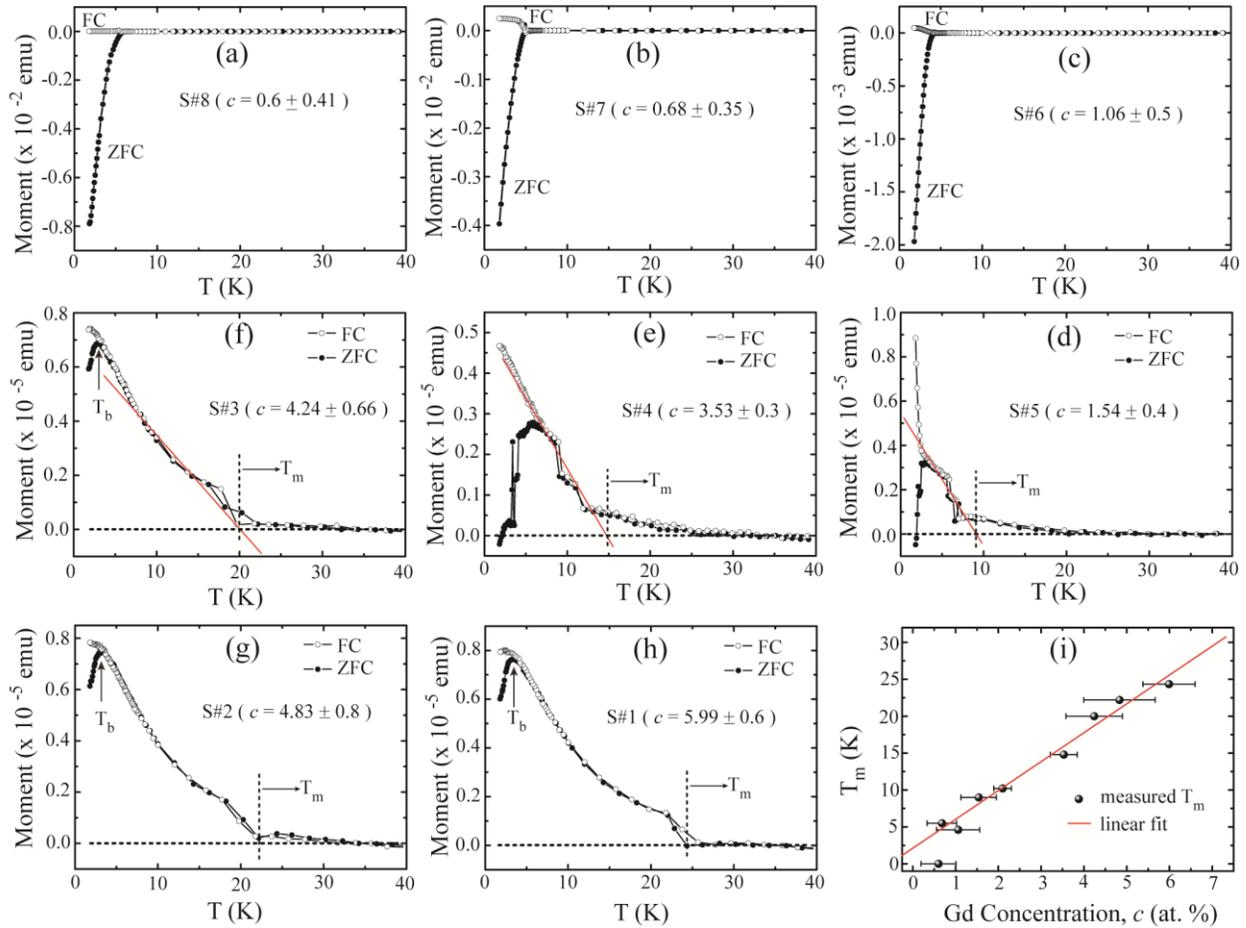

**Figure S4: DC M-T measurements for a set of samples grown at single sputtering run.** The transition from diamagnetic to paramagnetic state with increasing Gd concentration is evident from (a) to (h). The magnetic transition temperature $T_m$ is shown in (d)-(h). The red lines in (d)-(f) represent the slope of the positive magnetization and the intersect points of theses slopes to the zero-magnetization provide the values of $T_m$. With higher Gd incorporation the blocking temperature $T_b$ appears in the M-T measurements. (i) Variation of $T_m$ with Gd concentration. The spherical scattering points represent the measured $T_m$ values and the red line is a linear fit.

The DC M-T measurements in both FC and ZFC conditions for a set of samples prepared in a single sputtering run are shown in Figure S3. The sample S#8 was taken from the extreme right of a sample strip while the sample S#1 was placed at the extreme left (see Figure S1 for reference). The order of Gd concentration is decreasing from S#1 to S#2 to S#3 and so on up to S#8. The transition from a diamagnetic state to a paramagnetic state with increasing Gd is evident from the figure. The red lines in (d)-(f) of Figure S3 closely follow the magnetization curve in the paramagnetic state. The dotted black lines indicate the zero magnetization state. We have extracted the $T_m$ values from the intersect points of these red lines to the zero-magnetization line. The samples, with relatively higher Gd concentration shown in (g)-(h) of Figure S3, switches sharply to the positive magnetization state and the switching temperature is considered as the magnetic transition temperature $T_m$. Finally, we have plotted the $T_m$ values for this set of samples with Gd concentration in Figure S3 (i) and the dependence is linear as it is evident by the linear fit [the red line in Figure S3 (i)].

From the M-T data presented in Figure S4, we observe several interesting phenomena, namely, (i) the onset of positive magnetization in ZFC changes from a gradual increase [(d) & (e)] to a relatively sharp switching [(f) - (h)] for increasing c-values, (ii) at the peak ($T_b$) of ZFC, FC and ZFC starts to deviate, (iii) at c ~ 1%, the $T_m$ value deviates from the linear behaviour [Figure S4(i)]. The first two points can be understood in the light of superparamagnetism in the Gd particles. From the AFM studies we have got the indication that the particle size increases with Gd concentration. The average particle size for c = 5 % is about 50-60 nm [Figure 1(d) in the main text] and the same for c = 9 % is about 130-140 nm [Figure 1(e) in the main text]. Hence for the lower range of c- values Gd particles can be treated as superparamagnetic in nature. At relative low Gd concentration the thermal fluctuation can have stronger influence on the particles due to their smaller size and hence positive the magnetization starts to appear smoothly [(d) & (e)] whereas, the thermal fluctuation is supposed to have less impact on the larger particles and hence we observe a relatively sharp transition [(d) & (e)]. Also at $T_b$ the deviation of the ZFC and FC curves indicate of freezing of the magnetic moments implying the superparamagnetic nature of the Gd particles. In this case, $T_m$ behaves as the Curies temperature and the $T_b$ is related to the blocking temperature. The third point, the deviation in $T_m$ from the linear behaviour is not clear and it could be related to the influence of the superconducting Nb matrix on the magnetic ordering. However, we feel that we need to perform more experiments like the field dependent magnetization measurement along with their detailed structural analysis using high resolution TEM to have a clear understanding of the magnetic behaviour on the particle size and the clustering effect for high Gd concentration.